%% file: main.tex
\PassOptionsToPackage{table,dvipsnames}{xcolor}
\PassOptionsToPackage{numbers,sort&compress}{natbib}
\PassOptionsToPackage{bookmarks=false}{hyperref}
\documentclass[sigconf,screen,natbib,authorversion]{acmart}

\setcopyright{acmlicensed}
\acmPrice{15.00}
\acmDOI{10.1145/3611643.3613882}
\acmYear{2023}
\copyrightyear{2023}
\acmSubmissionID{fse23industry-p63-p}
\acmISBN{979-8-4007-0327-0/23/12}
\acmConference[ESEC/FSE '23]{Proceedings of the 31st ACM Joint European Software Engineering Conference and Symposium on the Foundations of Software Engineering}{December 3--9, 2023}{San Francisco, CA, USA}
\acmBooktitle{Proceedings of the 31st ACM Joint European Software Engineering Conference and Symposium on the Foundations of Software Engineering (ESEC/FSE '23), December 3--9, 2023, San Francisco, CA, USA}
\received{2023-05-18}
\received[accepted]{2023-07-31}

\input{src_packages.tex}

\input{src_commands.tex}

\input{src_acronyms.tex}

\begin{document}

\title[Automated Test Generation for Medical Rules Web Services: A Case Study at the Cancer Registry of Norway]{Automated Test Generation for Medical Rules Web Services:\\A Case Study at the Cancer Registry of Norway}

\author{Christoph Laaber}
\orcid{0000-0001-6817-331X}
\affiliation{%
  \institution{Simula Research Laboratory}
  \city{Oslo}
  \country{Norway}}
\email{laaber@simula.no}

\author{Tao Yue}
\orcid{0000-0003-3262-5577}
\affiliation{%
  \institution{Simula Research Laboratory}
  \city{Oslo}
  \country{Norway}}
\email{tao@simula.no}

\author{Shaukat Ali}
\orcid{0000-0002-9979-3519}
\affiliation{%
  \institution{Simula Research Laboratory and Oslo Metropolitan University}
  \city{Oslo}
  \country{Norway}}
\email{shaukat@simula.no}

\author{Thomas Schwitalla}
\orcid{0000-0002-0286-1686}
\affiliation{%
  \institution{Cancer Registry of Norway}
  \city{Oslo}
  \country{Norway}}
\email{thsc@kreftregisteret.no}

\author{Jan F. Nygård}
\orcid{0000-0001-9655-7003}
\affiliation{%
  \institution{Cancer Registry of Norway}
  \city{Oslo}
  \country{Norway}}
\affiliation{%
  \institution{UiT The Arctic University of Norway}
  \city{Tromsø}
  \country{Norway}}
\email{jfn@kreftregisteret.no}

\begin{abstract}
The \gls{crn} collects, curates, and manages data related to cancer patients in Norway, supported by an interactive, human-in-the-loop, socio-technical decision support software system.
Automated software testing of this software system is inevitable; however, currently, it is limited in \gls{crn}'s practice.
To this end, we present an industrial case study to evaluate an AI-based system-level testing tool, i.e., \evomaster{}, in terms of its effectiveness in testing \gls{crn}'s software system.
In particular, we focus on \guri{}, \gls{crn}'s medical rule engine, which is a key component at the \gls{crn}.
We test \guri{} with \evomaster{}'s black-box and white-box tools and study their test effectiveness regarding code coverage, errors found, and domain-specific rule coverage.
The results show that all \evomaster{} tools achieve a similar code coverage; i.e., around \num{19}\% line, \num{13}\% branch, and \num{20}\% method; and find a similar number of errors; i.e., \num{1} in \guri{}'s code.
Concerning domain-specific coverage, \evomaster{}'s black-box tool is the most effective in generating tests that lead to applied rules; i.e., \num{100}\% of the aggregation rules and between \num{12.86}\% and \num{25.81}\% of the validation rules; and to diverse rule execution results; i.e., \num{86.84}\% to \num{89.95}\% of the aggregation rules and \num{0.93}\% to \num{1.72}\% of the validation rules pass, and \num{1.70}\% to
\num{3.12}\% of the aggregation rules and \num{1.58}\% to \num{3.74}\% of the validation rules fail.
We further observe that the results are consistent across \num{10} versions of the rules.
Based on these results, we recommend using \evomaster{}'s black-box tool to test \guri{} since it provides good results and advances the current state of practice at the \gls{crn}. 
Nonetheless, \evomaster{} needs to be extended to employ domain-specific optimization objectives to improve test effectiveness further.
Finally, we conclude with lessons learned and potential research directions, which we believe are applicable in a general context.
\end{abstract}

\keywords{automated software testing, test generation, REST APIs, cancer registry, electronic health records, rule engine}

\ccsdesc[500]{General and reference~Validation}
\ccsdesc[500]{Software and its engineering~Software testing and debugging}
\ccsdesc[500]{Applied computing~Health care information systems}

\maketitle

\glsresetall{}

\input{src_introduction.tex}

\input{src_background.tex}

\input{src_study.tex}

\input{src_results.tex}

\input{src_discussion.tex}

\input{src_rw.tex}

\input{src_conclusions.tex}

\begin{acks}
This work received funding from \gls{rcn} under the project \texttt{309642} and has benefited from the \glsentryfull{ex3}, which is supported by the \gls{rcn} project \texttt{270053}.
\end{acks}

\bibliographystyle{ACM-Reference-Format}
\bibliography{refs.bib,refs-tweets.bib,refs-jan.bib}

\end{document}

%% file: src_packages.tex
\usepackage[T1]{fontenc}
\usepackage[utf8]{inputenc}
\usepackage[american]{babel}
\usepackage{csquotes}
\usepackage{microtype}
\usepackage{paralist}
\setdefaultenum{(1)}{(a)}{(i)}{A.}
\usepackage{enumitem}
\usepackage{url}
\usepackage{flushend}
\usepackage{etoolbox}
\usepackage{fix-cm}
\usepackage{textpos}
\usepackage{mfirstuc}
\usepackage{titlecaps}
\usepackage[datesep=.,style=ddmmyyyy]{datetime2}

\usepackage{algorithm}
\usepackage{algpseudocode}

\usepackage{amsmath}
\usepackage{amssymb}
\usepackage{bm}
\usepackage{relsize}
\usepackage{siunitx}
\usepackage[super]{nth}

\usepackage{booktabs}
\usepackage{multirow}
\usepackage{colortbl}
\usepackage{makecell}
\usepackage{rotating}
\usepackage{threeparttable}

\usepackage{float}
\usepackage{graphicx}
\usepackage{xcolor}

\usepackage{listingsutf8}

\usepackage{xparse}
\usepackage{xspace}

\usepackage[xindy,acronym]{glossaries}
\glsdisablehyper

\usepackage{natbib}

\usepackage{hyperref}
\usepackage[all]{hypcap}

\usepackage[capitalise]{cleveref}

%% file: src_commands.tex
\newcommand{\code}[1]{\texttt{#1}}
\newcommand{\tool}[1]{\textit{#1}}

\newcommand{\version}[1]{\texttt{#1}}

\newcommand{\evomaster}{\tool{EvoMaster}}
\newcommand{\evomasterbb}{\tool{\evomaster-BB}}
\newcommand{\evomasterwb}{\tool{\evomaster-WB}}
\newcommand{\evomasterwbmio}{\tool{\evomasterwb-MIO}}
\newcommand{\evomasterwbmosa}{\tool{\evomasterwb-MOSA}}
\newcommand{\evomasterwbwts}{\tool{\evomasterwb-WTS}}
\newcommand{\evosuite}{\tool{EvoSuite}}
\newcommand{\guri}{\tool{GURI}}
\newcommand{\numversions}{\num{10}}

\newcommand{\rqcodecov}{RQ~1}
\newcommand{\rqfaults}{RQ~2}
\newcommand{\rqexecutedrules}{RQ~3}
\newcommand{\rqruleresults}{RQ~4}

\newcommand{\nrcols}[2]{\multicolumn{#1}{l}{#2}}

\definecolor{greylight}{RGB}{240,240,240}
\definecolor{greymedium}{RGB}{189,189,189}
\definecolor{greydark}{RGB}{99,99,99}

\usepackage{tcolorbox}
\tcbset{colback=greylight, colframe=greydark, arc=1pt, boxrule=0.75pt}
\newcommand{\summarybox}[2]{
	\begin{tcolorbox}
		\small{
			\textbf{#1:} #2
		}
	\end{tcolorbox}
}

%% file: src_acronyms.tex
\newacronym{api}{API}{application programming interface}
\newacronym{caress}{CaReSS}{Cancer Registration Support System}
\newacronym{cpu}{CPU}{central processing unit}
\newacronym{crn}{CRN}{Cancer Registry of Norway}
\newacronym{db}{DB}{database}
\newacronym{ea}{EA}{evolutionary algorithm}
\newacronym{ehr}{EHR}{electronic health record}
\newacronym{ex3}{eX\textsuperscript{3}}{Experimental Infrastructure for Exploration of Exascale Computing}
\newacronym{ga}{GA}{genetic algorithm}
\newacronym{gan}{GAN}{generative adversarial network}
\newacronym{gui}{GUI}{graphical user interface}
\newacronym{hpc}{HPC}{high-performance computing}
\newacronym{html}{HTML}{HyperText Markup Language}
\newacronym{http}{HTTP}{Hypertext Transfer Protocol}
\newacronym{io}{I/O}{input-output}
\newacronym{jar}{JAR}{Java Archive}
\newacronym{jdk}{JDK}{Java Development Kit}
\newacronym{json}{JSON}{JavaScript Object Notation}
\newacronym{jvm}{JVM}{Java virtual machine}
\newacronym{llm}{LLM}{large language model}
\newacronym{loc}{LOC}{lines of code}
\newacronym{maoea}{MaOEA}{many-objective evolutionary algorithm}
\newacronym{mio}{MIO}{Many Independent Objective}
\newacronym{ml}{ML}{machine learning}
\newacronym{moea}{MOEA}{multi-objective evolutionary algorithm}
\newacronym{mosa}{MOSA}{Many-Objective Sorting Algorithm}
\newacronym{ocl}{OCL}{Object Constraint Language}
\newacronym{pp}{pp}{percentage points}
\newacronym{rcn}{RCN}{The Research Council of Norway}
\newacronym{rest}{REST}{representational state transfer}
\newacronym{rmit}{RMIT}{randomized multiple interleaved trials}
\newacronym{rpc}{RPC}{remote procedure call}
\newacronym{rq}{RQ}{research question}
\newacronym{sql}{SQL}{structured query language}
\newacronym{sut}{SUT}{subject under test}
\newacronym{uml}{UML}{Unified Modeling Language}
\newacronym{vae}{VAE}{variational autoencoder}
\newacronym{wts}{WTS}{Whole Test Suite}

%% file: src_introduction.tex
\section{Introduction}

Cancer is a leading cause of death worldwide, with nearly \num{10} million deaths in 2020~\citep{who:999}.
Consequently, most countries systematically collect data about cancer patients in specialized registries for the ultimate purpose of improving patient care, by supporting decision-making and conducting research.
These registries maintain specialized software systems to collect, curate, and analyze cancer data.
However, engineering such software systems poses many challenges, such as
\begin{inparaenum}
    \item collecting data for patients throughout their lives from diverse sources, e.g., hospitals, laboratories, and other registries;
    \item dealing with continuous evolution, e.g., due to software updates, new requirements, updated regulations, and new medical research;
    and
    \item increased incorporation of machine learning algorithms for decision support and production of statistics for relevant stakeholders including patients and policymakers.
\end{inparaenum}

Our context is the \gls{crn}, which has developed a \gls{caress} to support collecting, processing, and managing cancer-related data from various medical entities such as Norwegian hospitals and laboratories.
Based on the processing, the \gls{crn} generates statistics that are consumed by external entities, including policymakers, hospitals, and patients.
It further provides data for researchers to conduct research.
Naturally, the quality of statistics and data depends on how correct and reliable \gls{caress} is.
To this end, testing is one method to ensure the dependability of \gls{caress} to a certain extent.

This paper reports on a case study in the real-world context of the \gls{crn}, focusing on testing one of the key components of \gls{caress}, called \guri{}-- a rule engine responsible for checking medical rules for various purposes such as data validation and aggregation.
Currently, \guri{}'s testing is primarily manual, as also reported by \citet{haas:21} that manual testing is a common practice in industry.
Our study takes a popular open-source, AI-based system-level testing tool called \evomaster{}~\citep{arcuri:19}, which has been shown to be superior to \num{9} other \gls{rest} \gls{api} testing tools~\citep{kim:22}, and performs automated testing of \guri{} to assess its effectiveness in testing \guri{} from various perspectives.

We apply \evomaster{}'s four tools, i.e., black-box and white-box tools with three \glspl{ea}, to test \guri{}'s ten versions.
In particular, we assess \evomaster{}'s capability to achieve source code coverage, errors found, and domain-specific rule coverage.
The results of our experiments show that all four tools' effectiveness is similar in terms of code coverage and errors found across all \guri{} versions.
However, we observe that \evomaster{}'s black-box tool is more effective for domain-specific coverage.
We further compare the results with \guri{} in production, and the results of the black-box tool were closer to the production system.

Based on our results, we recommend using \evomaster{}'s black-box tool as the starting point to automate the testing of \guri{} and \gls{caress}.
However, this study also shows that \evomaster{} must be customized for \gls{crn}'s context by explicitly incorporating domain-specific coverage elements (e.g., coverage of different types of cancers and treatments) in the search process, e.g., encoding as new fitness functions, as also argued by \citet{boehme:22}.
Finally, we provide detailed discussions and lessons from our industrial case study regarding its generalization to other contexts and point out key research areas that deserve attention from the software engineering community.

%% file: src_background.tex
\section{Background and Context}
\label{sec:background}
In this section, we first discuss the \gls{crn} context, \guri{} and the medical rules that \guri{} relies on, followed by background on \evomaster{}. 

\subsection{Application Context}
The \gls{crn} regularly collects cancer patients' data (e.g., diagnostic, treatment), based on which cancer research and statistics can be conducted.
To ensure the quality of the collected data, the \gls{crn}'s \gls{caress} has introduced several preventive efforts to discover and amend inaccurate or missing data.
Patient data is submitted to the \gls{crn} as \textit{cancer messages}, from which \textit{cancer cases} are derived via a coding and aggregation process, representing a timeline of a patient's diagnoses, treatments, and follow-ups.
The coding process relies on standard classification systems and depends on hundreds of \textit{medical rules} for validating cancer messages and aggregating cancer cases.
Consequently, these rules are of two categories: \textit{validation rules} and \textit{aggregation rules}, which are defined by medical experts and implemented in \guri{} for automated validation and aggregation of cancer messages and cases.
These rules constantly evolve, e.g., due to updated medical knowledge and procedures.

Below is a validation rule, which states that for all the cancer messages of type \textit{H}, if the \textit{surgery} value is equal to \num{96}, then the \textit{basis} value must be greater than \num{32}. 
\begin{equation*}
	\forall \; messageType = H \implies (surgery = 96 \implies basis > 32)
\end{equation*}

The aggregation rule below determines the state of \emph{Morphologically verified}, based on a given \emph{Basis} value of cancer messages:
\vspace{-1em}
\begin{algorithm}[ht]
    \small
    \begin{algorithmic}[0]
   \If{Basis $\in$ ['22', '32', '33', '34', '35', '36', '37', '38', '39', '57', '60', '70', '72', '74', '75', '76', '79']}
   \State Morphologically verified = 'Yes'
    \ElsIf{Basis $\in$ ['00', '10', '20', '23', '29', '30', '31', '40', '45', '46', '47', '90', '98']}
    \State Morphologically verified = 'No'
    \Else 
    \State Morphologically verified = null
    \EndIf
\end{algorithmic}
\end{algorithm}

These rules are stored in \guri{}'s internal database.
New rules and updates to the rules are done through \gls{gui}, which is available for medical personnel.

\subsection{\evomaster{}}
\evomaster{}~\citep{arcuri:19} is an open-source, automated and search-based software testing framework.
\Gls{ea} are employed to optimize search objectives involving coverage criteria such as line, branch, and method coverages. 
\evomaster{} can be used with different programming languages, such as Java and Python.
The black-box testing tool of \evomaster{}~\citep{arcuri:21} relies on random testing with multi-objective search to maximize black-box metrics (e.g., endpoint coverage and status code coverage) and fault detection capability (\num{500} status code). 
\evomaster{} also defines a series of white box testing tools~\citep{zhang:22a, marculescu:22}, combined with different multi-objective \glspl{ea} (e.g., \gls{mio}), for achieving various test generation purposes by providing a comprehensive list of coverage criteria, including lines, branches, and faults.

%% file: src_study.tex
\section{Experimental Study}
\label{sec:study}

We perform a laboratory experiment~\citep{stol:18} of automated test generation techniques for \gls{rest} \glspl{api} at the \gls{crn} to understand their effectiveness of covering source code, revealing errors, and executing domain-specific elements, i.e., medical rules.
Our experiment comprises a real-world study subject, i.e., \gls{crn}['s] medical rule engine \guri{}, and four \gls{rest} \gls{api} test generation tools.

\subsection{Research Questions}
\label{sec:study:rqs}

Our study investigates the following four \glspl{rq} to assess the effectiveness of the test generation tools:

\begin{description}[topsep=0.8em,labelindent=1em,labelwidth=3em,itemsep=0.2em]
    \item[\rqcodecov] How much code coverage do the tools achieve?
    \item[\rqfaults] How many errors do the tools trigger?
    \item[\rqexecutedrules] How many rules can the tools execute?
    \item[\rqruleresults] Which results do the rule executions yield, and how do these compare to production \guri{} rule execution results?
\end{description}

\rqcodecov{} and \rqfaults{} are \enquote{traditionally} investigated research questions for evaluating test generation tools in terms of effectiveness, including \gls{rest} \gls{api} test generation~\citep{kim:22}. 
\rqexecutedrules{} and \rqruleresults{} are domain-specific \glspl{rq} and evaluate the test generation tools' effectiveness in testing \guri{}'s main functionality, i.e., validating and aggregating cancer messages and cancer cases.
\rqexecutedrules{} studies the test generation tools' capability to execute the medical rules with the ultimate goal of testing them.
The goal of \rqruleresults{} is to investigate which results the executed rules yield with test generation tools and how these results compare to the results from the production system.

\subsection{Study Subject --- \guri{}}
\label{sec:study:subject}

\guri{} is implemented as a Java web application with Spring Boot, which exposes \gls{rest} \gls{api} endpoints to \gls{crn}['s] internal systems and provides a web interface for medical coders.
\Cref{tab:study_subject} shows an overview of \guri{}.
Although \guri{} has \num{32} \gls{rest} endpoints, we only focus on \num{2} in this experiment, as these are the ones handling the rules, i.e., one for validating cancer messages with validation rules and one for aggregating cancer messages into cancer cases with aggregation rules.
\guri{}'s most recent version consists of \num{71} validation rules and \num{43} aggregation rules.
Since \guri{} was introduced, its source code has hardly changed; however, its rules have been subject to evolution due to updated medical knowledge, such as rule additions, deletions, and modifications.
Consequently, based on these changes, we form ten rule sets as ten versions in this experiment.
Specifically, out of \num{28} unique points in time where the rules were changed in \guri{}, we select \num{10} dates where the changes are most severe (the most rule additions and deletions occurred).
\Cref{tab:rule_versions} depicts this rule evolution.

\input{src_tab_guri_meta_data.tex}
\input{src_tab_rule_versions.tex}

\subsection{Test Generation Tools}
\label{sec:study:tools}

The automated \gls{rest} \gls{api} test generation tools form the independent variable of our experimental study.
We select \evomaster{} in version \version{v1.5.0}\footnote{\url{https://github.com/EMResearch/EvoMaster/releases/tag/v1.5.0}} with multiple parameterizations as the tools~\citep{arcuri:19}.
\evomaster{} was recently shown to be the most effective tool, in terms of source code coverage and triggered errors, among ten different tools~\citep{kim:22}.
A tool, in the context of our study, is a specific \evomaster{} parameterization, consisting of the testing approach and the employed \gls{ea}.

In terms of testing approach, our experiments use both black-box and white-box testing.
\evomasterbb{} relies on a random-testing approach to generate the \gls{rest} requests~\citep{arcuri:21}.
On the other hand, \evomasterwb{} uses an \gls{ea} to generate tests based on randomly generated (initial) \gls{rest} requests, coverage feedback, and mutation~\citep{arcuri:19}.
\evomaster{} supports three \glspl{ea}:
\begin{inparaenum}
    \item \gls{mio}~\citep{arcuri:18b}, which is a \gls{maoea} focusing on scalability in the presence of many testing targets, that was specifically designed for \gls{rest} \gls{api} test generation and is \evomaster{}'s default;
    \item \gls{mosa}~\citep{panichella:18}, which is the first \gls{maoea} that was designed for unit test generation with \evosuite{}~\citep{fraser:11};
    and
    \item \gls{wts}~\citep{fraser:13}, which is a single objective \gls{ga} that was designed for unit test generation and is the original \gls{ea} of \evosuite{}.
\end{inparaenum}
Our experiment investigates all four \evomaster{} tool parameterizations:
\begin{inparaenum}
    \item \textbf{\evomasterbb{}:} \evomaster{} black-box;
    \item \textbf{\evomasterwbmio{}:} \evomaster{} white-box with \gls{mio};
    \item \textbf{\evomasterwbmosa{}:} \evomaster{} white-box with \gls{mosa};
    and
    \item \textbf{\evomasterwbwts{}:} \evomaster{} white-box with \gls{wts}.
\end{inparaenum}
Beyond the testing approach and the \gls{ea}, our experiment uses the default parameters of \evomaster{}, similar to \citet{kim:22}.

\subsection{Evaluation Metrics}
\label{sec:study:metrics}
To evaluate the effectiveness of the tools and answer the \glspl{rq}, we rely on the following evaluation metrics, which are the dependent variables of our study.

\textbf{\rqcodecov{} - code coverage.} We rely on \textbf{line}, \textbf{branch}, and \textbf{method coverage} extracted with \tool{JaCoCo}\footnote{version \version{0.8.8} available at \url{https://www.jacoco.org/jacoco}}, similar to \citet{kim:22}.

\textbf{\rqfaults{} - errors.} We use three types of \num{500} errors triggered by the tools, as defined by \citet{kim:22}:
\begin{inparaenum}
    \item \textbf{Unique Errors} are the number of errors grouped by their complete stack traces;
    \item \textbf{Unique Failure Points} are the number of occurrences of the same error, i.e., the first line of a stack trace;
    and
    \item \textbf{Unique Library Failure Points} are the number of errors that are unique failure points occurring in the library code.
\end{inparaenum}

\textbf{\rqexecutedrules{} - rule execution status.} The first domain-specific metric is the number of executed rules during a tool's execution.
We distinguish three types:
\begin{inparaenum}
    \item \textbf{Applied} is a rule that is fully executed at least once on an input, irrespective of the result (pass, fail, or warning);
    \item \textbf{Not Applied} is a rule that is partially executed against an input, which only applies to validation rules of the form $apply \implies rule$, where $apply$ is the condition on which $rule$ is applied;
    \item \textbf{Not Executed} is a rule that is never executed on any input.
\end{inparaenum}

\textbf{\rqruleresults{} - rule execution results.} The second domain-specific metric are the execution results of rules that have been applied (see \rqexecutedrules{}).
We distinguish three types:
\begin{inparaenum}
    \item \textbf{Pass} results from a successful rule execution, i.e., the cancer message with all its data is valid, or the cancer case was successfully aggregated with a previous cancer case and a number of cancer messages; 
    \item \textbf{Fail} results from an unsuccessful rule execution, i.e., the cancer message is invalid, or the cancer case fails to be aggregated;
    \item \textbf{Warning} is the result where the rule execution is successful; however, the data appears to be dubious (e.g., a patient's age is \num{120} years, which is theoretically valid but highly unlikely). 
\end{inparaenum}

To answer \rqruleresults{}, we compare the results of the tools, based on these three categories, to the results of the production \guri{}, deployed at the \gls{crn} to determine if their similarity.

\subsection{Experiment Setup}
\label{sec:study:setup}

The experiment setup is concerned with the experiment execution settings and execution environment.

To deal with the stochastic nature of the \glspl{ea} underlying \evomaster{}, each tool is executed repeatedly for \num{30} repetitions~\citep{arcuri:11}.
The analyses then rely on the arithmetic mean across all the repetitions.
Each tool is executed for \num{1} hour for each repetition and version by following the practice of \citet{kim:22}.
They also observed that for complex and constrained input parameters, source code coverage and thrown errors hardly increase after \num{10} minutes, which is also our case.
Beyond these settings, the experiments consider the tool parameterizations from \cref{sec:study:tools}.
The experiment further randomizes the order of the tools and versions in each repetition, with \gls{rmit}~\citep{abedi:17}, reducing potential confounding factors that stem from the execution environment or the execution order.

We executed the experiments on the \gls{ex3} \gls{hpc} cluster\footnote{\url{https://www.ex3.simula.no}} hosted at the first author's institution, which uses Slurm \version{20.02.7} as its cluster management software.
The experiments were scheduled on nodes of the same type (using the \tool{slowq} partition of \gls{ex3}), with the whole node exclusively reserved.
The nodes have \num{8} Intel(R) Xeon(R) Silver 4112 \glspl{cpu} @ \SI{2.60}{GHz} each with \num{4} cores, run Ubuntu \version{18.04.1}, and have \SI{40}{GB} total memory.
The experiments were conducted in the first half of 2023.
The Java tools were exclusively built and ran with OpenJDK \version{11.0.18} built by Adoptium\footnote{\url{https://adoptium.net}}.

\subsection{Threats to Validity}
\label{sec:study:threats}

We classify threats into construct, internal, and external validity.

The biggest threat to the \textbf{construct validity} concerns the choice of the metrics for evaluating the tools' effectiveness.
We rely on metrics widely used in the \gls{api} test generation research, in particular, adapted by a recent publication~\citep{kim:22}, to answer \rqcodecov{} and \rqfaults{}.
For \rqexecutedrules{} and \rqruleresults{}, we employ two sets of domain-specific metrics, i.e., executed rules in \rqexecutedrules{} and rule execution results in \rqruleresults{}, which are of specific interest to the \gls{crn}.
Nevertheless, it is unclear whether these domain-specific metrics are correlated with \enquote{good} test cases for the domain experts or sufficiently targeted to assess test generation tools for the \gls{crn}.
Further investigation via dedicated empirical studies is required to answer this question. 

In terms of \textbf{internal validity}, a crucial threat is that \evomaster{}white-box requires manually creating a \gls{sut} driver.
Failure to do so concerning how \evomaster{} expects the driver to be implemented and \guri{} requires to be controlled could threaten the \evomasterwb{} results.
Similarly, an incorrect injection of the \tool{JaCoCo} code coverage agent, implementation of the analyses scripts, and adaptation of \guri{} to retrieve rule executions could alter the study's results and implications.
We thoroughly tested our implementations to validate the correct behavior to mitigate this threat. 
Further internal validity threats relate to the experiment design include:
\begin{inparaenum}
    \item the number of repetitions (\num{30} in our study),
    \item the time budget for test generation (\num{1} hour),
    and
    \item the tool parameters (see \cref{sec:study:tools}).
\end{inparaenum}
These design decisions are based on previous research~\citep{arcuri:11, kim:22};
however, different experiment design decisions might lead to different results.
Finally, \evomaster{} relies on \tool{OpenAPI}\footnote{\url{https://www.openapis.org}} schema definitions to generate tests.
An incorrect schema definition potentially leads to sub-optimal tests, which could impact the reported results.
Our experiment uses the schema definitions generated by \tool{springdoc-openapi}\footnote{\url{https://springdoc.org}}, which \guri{} already employs.
Considering many parameters to configure, we try to build the base of our empirical study on the knowledge built by the existing literature to mitigate these internal threats. 

The primary \textbf{external validity threat} is related to the generalization is inherent to the study design: a case study on a single study subject, i.e., \gls{crn}['s] rule engine \guri{}.
All results are only valid in the context of our case study and are probably not transferable to other case studies.
Nevertheless, we provide implications and \enquote{more general} conclusions in the discussion section.
Moreover, our results are tightly coupled with the test generation tool(s), i.e., \evomaster{} in its four parameterizations (see \cref{sec:study:tools}), and do not generalize to other \gls{rest} \gls{api} test generation tools.
Finally, we perform a laboratory experiment on the research \gls{hpc} cluster \gls{ex3} with a standalone version of \guri{} and not a field experiment against the real-world \guri{} (or the testing environment) hosted at the \gls{crn}.
Consequently, our results might not generalize to the real system.
The extraction of the real-world \guri{} into a standalone version for the empirical study, was, however, done by the \gls{crn} developers, which should reduce this threat.

%% file: src_tab_guri_meta_data.tex
\begin{table}
    \centering
    \caption{\guri{} Meta Data}
    \label{tab:study_subject}
    
    \begin{tabular}{llr}
        \toprule
        
        Metric & & Value \\
        
        \midrule

        Rules
        & Validation & \num{71} \\
        & Aggregation & \num{43} \\
        
        \Gls{rest} Endpoints
        & Rule Handling & \num{2} \\
        & Total & \num{32} \\
        
        Versions
        & Selected & \numversions{} \\
        & Total & \num{28} \\
        
        \bottomrule 
    \end{tabular}
\end{table}

%% file: src_tab_rule_versions.tex
\begin{table}
    \centering
    \caption{Rule Evolution}
    \label{tab:rule_versions}
    \include{src_tab_rule_versions_body.tex}
\end{table}

%% file: src_tab_rule_versions_body.tex
\begin{tabular}{llrr}
\toprule
Version & Date & \nrcols{2}{Rules} \\
\cmidrule(l{1pt}r{1pt}){3-4}
 &  & Validation & Aggregation \\
\midrule
\version{v1} & \DTMdate{2017-12-12} & \num{30} & \num{32} \\
\version{v2} & \DTMdate{2018-05-30} & \num{31} & \num{33} \\
\version{v3} & \DTMdate{2019-02-06} & \num{48} & \num{35} \\
\version{v4} & \DTMdate{2019-08-27} & \num{49} & \num{35} \\
\version{v5} & \DTMdate{2019-11-11} & \num{53} & \num{37} \\
\version{v6} & \DTMdate{2020-09-25} & \num{56} & \num{37} \\
\version{v7} & \DTMdate{2020-11-24} & \num{66} & \num{38} \\
\version{v8} & \DTMdate{2021-04-20} & \num{69} & \num{43} \\
\version{v9} & \DTMdate{2022-01-13} & \num{69} & \num{43} \\
\version{v10} & \DTMdate{2022-01-21} & \num{70} & \num{43} \\
\bottomrule
\end{tabular}

%% file: src_results.tex
\section{Results}
\label{sec:results}
This section presents the results for each \gls{rq}.

\subsection{\rqcodecov{}: Code Coverage}
\label{sec:results:code-coverage}

\input{src_tab_coverages.tex}

We study the code coverage achieved by the tools in terms of line, branch, and method coverages.
\Cref{tab:coverages} depicts the coverage results.
Each coverage value consists of the arithmetic mean and standard deviation across all the versions and repetitions.
We refrain from reporting coverages for each version, as the source code does not change between versions; only the rules change (see \cref{sec:study:subject}).

We observe that all tools perform similarly, i.e., approximately \num{20}\% line coverage, \num{14}\% branch coverage, and \num{20}\% method coverage.
Note that the relatively low (absolute) coverage values must be considered with caution because we generated tests for (only) \num{2} of \num{32} \gls{rest} endpoints (see \cref{sec:study:subject}), as only these two handle the medical rules.
However, the code coverage is still of interest as a relative comparison among the tools.
With regards to it, the tools perform similarly, which is a different observation as obtained by \citet{kim:22}, where \evomasterwbmio{} achieves a higher code coverage than \evomasterbb{} by \num{7.35} \gls{pp} (line), \num{7.87} \gls{pp} (branch), and \num{5.69} \gls{pp} (method).
In addition, \citet{kim:22} report that the line and method coverages are similarly higher than the branch coverage, which aligns with our findings.

A potential reason for the low coverage values is that \evomaster{} produces many invalid requests, as \guri{} expects specific \gls{json} input parameter values, i.e., medical variables contained in cancer messages and cases.
In particular, \evomaster{} often fails to provide valid date strings.
This behavior is also observed by \citet{kim:22}.
An alternative reason is that all the tools reach the maximum code coverage achievable through the two \gls{rest} \glspl{api} used in our experiment.
The remaining \glspl{rq} will shed more light on these hypotheses.

\summarybox{\rqcodecov{} Summary}{
All tools achieve a similar line, branch, and method coverage.
In the context of the \gls{crn}, opting for a simpler test generation tool, i.e., \evomasterbb{}, is preferred over a more complex tool, i.e., any \evomasterwb{}, to cover more source code.
}

\subsection{\rqfaults{}: Errors}
\label{sec:results:errors}

The second \gls{rq} studies the thrown errors by the tools in terms of unique \num{500} errors, unique failure points, and unique library failure points.
\Cref{tab:errors} shows the results for each tool across all the versions and repetitions.
Similar to \rqcodecov{}, we refrain from reporting errors for each version.
For each error category, the table depicts four values:
\begin{inparaenum}
    \item the total number of errors (\enquote{All});
    \item errors that are due to the tools' intervention, e.g., the attached Java agent (\enquote{Tool});
    \item \gls{io} errors that occur during the test generation (\enquote{\gls{io}});
    and
    \item remaining errors that are not attributed to the Tool and \gls{io} categories but actually due to errors thrown by the application, i.e., \guri{} (\enquote{Remaining}).
\end{inparaenum}

\input{src_tab_errors.tex}

Considering unique errors (i.e., errors with identical stack traces, see \cref{sec:study:metrics}), we observe that the different tools trigger between \num{3} and \num{6.46} errors (\enquote{All});
however, on a closer inspection, we notice that the three \evomasterwb{} tools experience a high number of tool-related errors, i.e., the coverage agent attached to the \gls{jvm} throws a \code{KillSwitch} exception when concurrent threads are running after the test generation has finished.
This scenario inflates the \enquote{All} errors.
Moreover, we can see that all tools suffer from a varying degree of \gls{io} errors, most often due to broken \gls{io} pipe exceptions.
Finally, the number of \enquote{Remaining} unique errors is similar across all four tools, i.e., approximately \num{1}.
These results align with the code coverage results from \rqcodecov{}, i.e., no tool is superior over the others.

Regarding the unique failure points (i.e., errors where the top-most line of the stack trace is identical, see \cref{sec:study:metrics}), we observe a similar trend:
\enquote{All} errors are inflated by tool-related and \gls{io} errors, leaving approximately \num{1} \enquote{Remaining} unique failure point, which is, again, consistent across all tools.
In most cases, this \num{1} failure point is caused by a date parsing exception, e.g., when the diagnosis date of a cancer message is malformed.
Although seldom, all four tools trigger a unique library failure point (i.e., a unique failure point where the cause is located in \guri{}'s source code, see \cref{sec:study:metrics}).
Upon closer inspection, we identify \num{1} \guri{} rule parsing error on specific inputs.

Compared to \citet{kim:22}, the tools reveal fewer errors, and no tool is superior in the context of the \gls{crn}.

\summarybox{\rqfaults{} Summary}{
    All the tools reveal the same number of errors and failure points.
    Similar to \rqcodecov{}, this suggests employing the most straightforward tool, i.e., \evomasterbb{}, for its error-revealing capabilities.
}

\subsection{\rqexecutedrules{}: Rule Execution Status}
\label{sec:results:executed-rules}

\input{src_fig_executed_rules.tex}

This domain-specific RQ evaluates the tools' ability to execute medical rules.
\Cref{fig:executed-rules} shows the number of (distinct) rules (on the y-axis) by each tool (on the x-axis) for each version of \guri{}.
The dashed line depicts the total number of rules in the particular version.
Each bar represents the arithmetic mean of applied, not applied, or not executed rules (see \cref{sec:study:metrics}), with the error bars indicating the standard deviations.
The first two rows depict the validation rules, whereas the last two rows show the aggregation rules.

We make three main observations:
\begin{inparaenum}
    \item there is a difference in effectiveness depending on which tool is employed, which is different from what we observe for \rqcodecov{} and \rqfaults{};
    \item the tools are not equally effective for validation and aggregation rules;
    and
    \item the tool effectiveness does not change across the ten versions.
\end{inparaenum}
We discuss these observations in detail below:

\subsubsection{Observation 1: Tool Differences}
\evomasterbb{} executes the most validation and aggregation rules, as the not executed category has near zero rules. 
\evomasterbb{} generate tests where \emph{all} the aggregation rules (between \num{32} and \num{43}) are applied (in all versions) and between \num{7.0000000} (\num{12.8571429}\%) and \num{9.666667} (\num{25.806452}\%) validation rules, depending on the version.
Whereas \evomasterwbmio{}, the best-performing white-box tool, only generates tests for which between \num{26.9000000} (\num{63.5658915}\%) and \num{34.600000} (\num{93.333333}\%) aggregation and between \num{6.1333333} (\num{9.5238095}\%) and \num{7.466667} (\num{24.086022}\%) validation rules are applied.

This is a surprising result because \evomasterwb{}, compared to \evomasterbb{}, is on par in our study (i.e., \rqcodecov{} and \rqfaults{}) and superior in a recent comparison of \gls{rest} \gls{api} test generation tools~\citep{kim:22}.
One reason is that the \gls{ea} of \evomasterwb{} optimizes for \enquote{irrelevant} objectives.
Once \evomasterwb{} finds better solutions for the source code coverage, it steers itself into a situation where it does not execute more (different) rules anymore.
Conversely, \evomasterbb{}, with its simpler approach (concerning covering source code), exercises more diverse inputs (i.e., cancer messages and cases), which leads to more executed rules.
This is particularly evident for aggregation rules, where \evomasterbb{} applies all the rules, followed by \evomasterwbmio{} which applies between \num{63.5658915}\% and \num{93.333333}\% of the rules, depending on the version.

Moreover, we notice that \evomasterbb{} has no variance in the number of executed rules among identical repetitions, whereas all the \evomasterwb{} experience a variance to a varying degree, as indicated by the error bars in \cref{fig:executed-rules}.
This means that the \evomasterwb{} tools cannot consistently execute the same rules for identical repetitions.

\subsubsection{Observation 2: Rule Type Differences}
This leads to the second observation, where the tools are not equally-effective in executing validation rules as they are for aggregation rules.
\evomasterbb{} executes all the aggregation rules but only achieves applying between \num{12.8571429}\% and \num{25.806452}\% of the validation rules.
It is worse for the \evomasterwb{} tools.
We conclude that \evomaster{} struggles to generate tests that cover considerably more validation rules.
The reason is inherent to many validation rules, which are only applied if the left part of an implication is true (see \cref{sec:study:metrics}).
\citet{kim:22} observe a similar situation, where the tools generate many invalid requests (due to invalid parameter values) that are rejected by the \glspl{api};
however, in our case, the requests are not rejected, as they conform with the \tool{OpenAPI}, but lead to rules that fall into the not applied category.

We further observe that the \evomasterwb{} tools suffer to a varying degree from rules not being executed, which is more pronounced for \evomasterwbmosa{} and \evomasterwbwts{} than for \evomasterwbmio{}.
This means that the tests never reach a point (in the source code) where the rules are considered for execution.
\citet{arcuri:18b} also finds that \gls{mosa} and \gls{wts} perform inferior to \gls{mio} on most (but not all) problem types for the source code coverage,
which has the consequence that more rules are not executed.

\subsubsection{Observation 3: Version Differences}
Finally, we observe that, with changing rule sets due to the addition, deletion, and modification of rules, the tools' effectiveness is hardly impacted.
For \evomasterbb{}, the standard deviation of the relative number of applied rules (with respect to the total number of rules) in each version is \num{0}\% and \num{4.4153747}\% for aggregation and validation rules, respectively.
There is slightly more variation for \evomasterwb{}, i.e., up to \num{10.8906612}\% standard deviation, suggesting that the rule evolution at the \gls{crn} is not a factor for choosing a particular tool over the other.

\summarybox{\rqexecutedrules{} Summary}{
    \evomasterbb{} is more effective in generating tests that apply the rules.
    In particular, aggregation rules are considerably easier to cover than validation rules.
    The \evomasterwb{} tools are less effective than \evomasterbb{}, and \evomasterwbmio{} performs the best among all the three white-box tools.
    All the tools are similarly effective in generating tests that lead to rules being applied across the rule set versions.
}

\subsection{\rqruleresults{}: Rule Execution Results}
\label{sec:results:rule-results}

This domain-specific \gls{rq} evaluates the tools' capabilities to generate tests that lead to the three rule execution results, i.e., pass, fail, and warning (see \cref{sec:study:metrics}).
\Cref{fig:rule-execution-results} shows the rule execution results for the applied rules relative to the total number of rules in a version (on the y-axis) per tool (on the x-axis), version and rule type.
For this \gls{rq}, not applied and not executed rules are disregarded.
Each bar represents the arithmetic mean, and the error bars are the standard deviations.
The first two rows are for the validation rules, and the last two rows are for the aggregation rules.

We make five observations:
\begin{inparaenum}
    \item there is a high degree of variance among repetitions in terms of the execution results;
    \item there is a difference in the prevalence of the individual execution result for aggregation and validation rules;
    \item the tools follow the same effectiveness ranking as in \rqexecutedrules{};
    \item the tools are differently effective (although minor) for different versions; and
    \item the result distribution of the generated tests is different from production \guri{}.
\end{inparaenum}

\input{src_fig_rule_execution_results.tex}

\subsubsection{Observation 1: Repetition Differences}
The first observation is that there is a high degree of variance for each execution result among the repetitions, irrespective of the tool, version, and rule type.
This means that the generated tests yield different rule execution results in each repetition.
This is caused by how \evomaster{} generates (new) variable values of cancer messages and cancer cases: \emph{randomly}. An effective test strategy should generate valid instead of random variable values, which leads to more rules being applied.

\subsubsection{Observation 2: Tool Differences}
\evomasterbb{}, again, is the tool that achieves the highest number of passes for both rule types.
It reaches between \num{86.835708179}\% and \num{89.94682487}\% and between \num{0.930049612}\% and \num{1.71530439}\% for aggregation and validation rules, respectively.
The best performing white-box tool overall is, again, \evomasterwbmio{} which passes for \num{57.248028648}\% to \num{81.78646039}\% (aggregation) and \num{0.073702601}\% to \num{0.39468975}\% (validation) of the rules.
Both \evomasterwbmosa{} and \evomasterwbwts{} are inferior to \evomasterwbmio{} in terms of passes.

In terms of fails, the results are not as clear:
\evomasterwbmio{} generates tests that can fail more aggregation rules, i.e., between \num{1.979988886}\% and \num{3.74879966}\%, than \evomasterbb{}, which fails for \num{1.695018702}\% to \num{3.11905671}\%.
However, \evomasterbb{} fails considerably more validation rules (\num{1.579910436}\% to \num{3.74060333}\%) than \evomasterwbmio{} (\num{0.232663353}\% to \num{1.26591665}\%) and \evomasterwbmosa{} (\num{0.020375048}\% to \num{1.52915063}\%).
\evomasterwbmio{} is better than \evomasterwbmosa{} for some versions, whereas the opposite is true for other versions.

In terms of warnings, \evomasterbb{} yields more than all the \evomasterwb{} techniques with between \num{7.647632920}\% and \num{10.11112335}\% for aggregation rules, again followed by \evomasterwbmio{} with \num{3.794059904}\% to \num{7.79807329}\%.
However, none of the techniques can execute warnings for validation rules.

\subsubsection{Observation 3: Rule Execution Result Differences}
We observe that depending on the rule type, the tests favor different execution results across all the tools and versions.
For validation rules, fails are more common than passes (not considering warnings, as none are thrown).
This is natural due to the random variable value generation of \evomaster{}, which is the same reason why there is a high variance among identical repetitions.
Nevertheless, the tools can still generate tests that pass many of the applied rules.
For aggregation rules, we notice a different behavior, i.e., most of the rules pass, followed by warnings and fails.

\subsubsection{Observation 4: Version Differences}
Similar to \rqexecutedrules{}, we do not observe big differences across different versions for \evomasterbb{}.
The standard deviation of the rules that pass is \num{1.00341804} and \num{0.29937054}, fail is \num{0.41151320} and \num{0.65759635}, and warning is \num{0.86034008} and \num{0} for aggregation and validation rules, respectively.
The \evomasterwb{} tools only exhibit more variance across the different versions for aggregation rules that pass and yield warnings.
Nevertheless, this strengthens our suggestion from \rqexecutedrules{} that rule evolution is not a deciding factor for choosing one tool over another.

\subsubsection{Observation 5: Comparison to Production \guri{}}

\input{src_tab_rule_execution_percentages.tex}

Finally, we compare the rule execution results of the tools to those from production \guri{}, i.e., to real-world statistics of rule execution results.
\Cref{tab:rule_execution_precentages} depicts this for \enquote{Production} and compares it to all the four tools.
The values are arithmetic means and standard deviations relative to the total number of rules.
Note that we only consider the tool results for the latest version, i.e., \version{v10}, as the production results are only available for the current version.

We observe that the distributions of the tools are considerably different from production \guri{}.
In production, most rules are not applied (\num{73.75}\%), and of the applied ones, the vast majority passes (\num{26.19}\%).
Only a fraction fails, and even fewer yield a warning.
The situation is even more extreme for aggregation rules: \num{99.88}\% of the rules pass, and only a negligible number fail or yield a warning.
The closest tool to production, in terms of effectiveness, is \evomasterbb{}.
However, none of the tools achieves a similar number of rules that pass for both rule types.

Interestingly, the tools often achieve higher failure and warning rates, as observed in production.
In particular, \evomasterbb{} (\num{1.61}\%) and \evomasterwbmio{} (\num{1.41}\%) can fail more validation rules,
and all four tools yield more failures and warnings for aggregation rules (\num{1.92}\% to \num{3.86}\%).

We conclude that in terms of passing rules, \evomasterbb{} performs best, but there is much room for improvement to reach production \guri{} levels.
In terms of failing and executing warnings, \evomasterbb{} performs best for validation and \evomasterwbwts{} for aggregation rules.

\summarybox{\rqruleresults{} Summary}{
    \evomasterbb{} is the tool that yields the most passes, fails, and warnings among all the tools, except for failing aggregation rules where \evomasterwbmio{} is the preferred tool.
    For all the tools, the validation rules are the easiest to fail, and aggregation rules are the easiest to pass.
    Warnings can only be triggered for aggregation rules. Compared to production \guri{}, no tool can pass a similar amount of rules, but all are good at generating tests that lead to fails and warnings;
    \evomasterbb{} is the closest to production.
}

%% file: src_tab_coverages.tex
\begin{table}
    \centering
    \caption{
    Source code coverage per tool across all the versions and repetitions.
    The values are arithmetic means $\pm$ standard deviations in percentages.
    }
    \label{tab:coverages}
    
    \include{src_tab_coverages_body.tex}
\end{table}

%% file: src_tab_coverages_body.tex
\begin{tabular}{llll}
\toprule
Tool & Line & Branch & Method \\
\midrule
\evomasterbb{} & \num{19.7366666666667}\%$\pm$\num{0.441177607273331} & \num{14.2433333333333}\%$\pm$\num{0.631494131682992} & \num{20.0333333333333}\%$\pm$\num{0.179805419417013} \\
\evomasterwbmio{} & \num{18.6833333333333}\%$\pm$\num{1.66502148455679} & \num{12.97}\%$\pm$\num{1.88350523578888} & \num{19.45}\%$\pm$\num{1.22985438407157} \\
\evomasterwbmosa{} & \num{18.14}\%$\pm$\num{1.53675155805048} & \num{12.37}\%$\pm$\num{1.77301398621605} & \num{19.1266666666667}\%$\pm$\num{1.02996357805959} \\
\evomasterwbwts{} & \num{19.2466666666667}\%$\pm$\num{2.25894481143133} & \num{13.76}\%$\pm$\num{2.30509106459018} & \num{19.68}\%$\pm$\num{1.91742581059357} \\
\bottomrule
\end{tabular}

%% file: src_tab_errors.tex
\begin{table*}
    \centering
    \footnotesize
    \caption{
    500 errors per tool across all the versions and repetitions.
    The values are arithmetic means $\pm$ standard deviations.
    }
    \label{tab:errors}
    
    \include{src_tab_errors_body.tex}
\end{table*}

%% file: src_tab_errors_body.tex
\begin{tabular}{llllllllll}
\toprule
Tool & \nrcols{4}{Unique Errors} & \nrcols{4}{Unique Failure Points} & Unique Library Failure Points \\
\cmidrule(l{1pt}r{1pt}){2-5}
\cmidrule(l{1pt}r{1pt}){6-9}
 & All & Tool & \Gls{io} & Remaining & All & Tool & \Gls{io} & Remaining &  \\
\midrule
\evomasterbb{} & \num{3}$\pm$\num{1.02477667034348} & \num{0}$\pm$\num{0} & \num{1.93}$\pm$\num{0.99081062046121} & \num{1.07}$\pm$\num{0.334873809407998} & \num{2.03333333333333}$\pm$\num{0.269190951029083} & \num{0}$\pm$\num{0} & \num{0.966666666666667}$\pm$\num{0.179805419417013} & \num{1.06666666666667}$\pm$\num{0.320256307610175} & \num{0.00333333333333334}$\pm$\num{0.0577350269189624} \\
\evomasterwbmio{} & \num{6.46333333333333}$\pm$\num{2.81574447444605} & \num{4.43333333333333}$\pm$\num{2.31590882112865} & \num{1.00666666666667}$\pm$\num{0.693945789008403} & \num{1.02333333333333}$\pm$\num{0.190376028769351} & \num{2.88}$\pm$\num{0.891580836469619} & \num{1.06333333333333}$\pm$\num{0.577726679271461} & \num{0.8}$\pm$\num{0.400668337976506} & \num{1.01666666666667}$\pm$\num{0.12823299585641} & \num{0.0166666666666667}$\pm$\num{0.12823299585641} \\
\evomasterwbmosa{} & \num{4.16333333333333}$\pm$\num{2.39355088554527} & \num{2.92}$\pm$\num{2.11546305722607} & \num{0.2}$\pm$\num{0.483852986323928} & \num{1.04333333333333}$\pm$\num{0.308392458701847} & \num{2.17}$\pm$\num{0.818535277187243} & \num{0.966666666666667}$\pm$\num{0.571528100903735} & \num{0.173333333333333}$\pm$\num{0.37916766204679} & \num{1.03}$\pm$\num{0.206337710649605} & \num{0.0233333333333333}$\pm$\num{0.151212122227632} \\
\evomasterwbwts{} & \num{6.04}$\pm$\num{1.74601313652266} & \num{4.78333333333333}$\pm$\num{1.53326183731332} & \num{0.18}$\pm$\num{0.384829371910397} & \num{1.07666666666667}$\pm$\num{0.52124326120029} & \num{2.5}$\pm$\num{0.64657575013984} & \num{1.28666666666667}$\pm$\num{0.47459423495064} & \num{0.18}$\pm$\num{0.384829371910397} & \num{1.03333333333333}$\pm$\num{0.197532154502578} & \num{0.0266666666666667}$\pm$\num{0.161376464931129} \\
\bottomrule
\end{tabular}

%% file: src_fig_executed_rules.tex
\begin{figure*}[tbp]
    \centering
    \includegraphics[width=\textwidth]{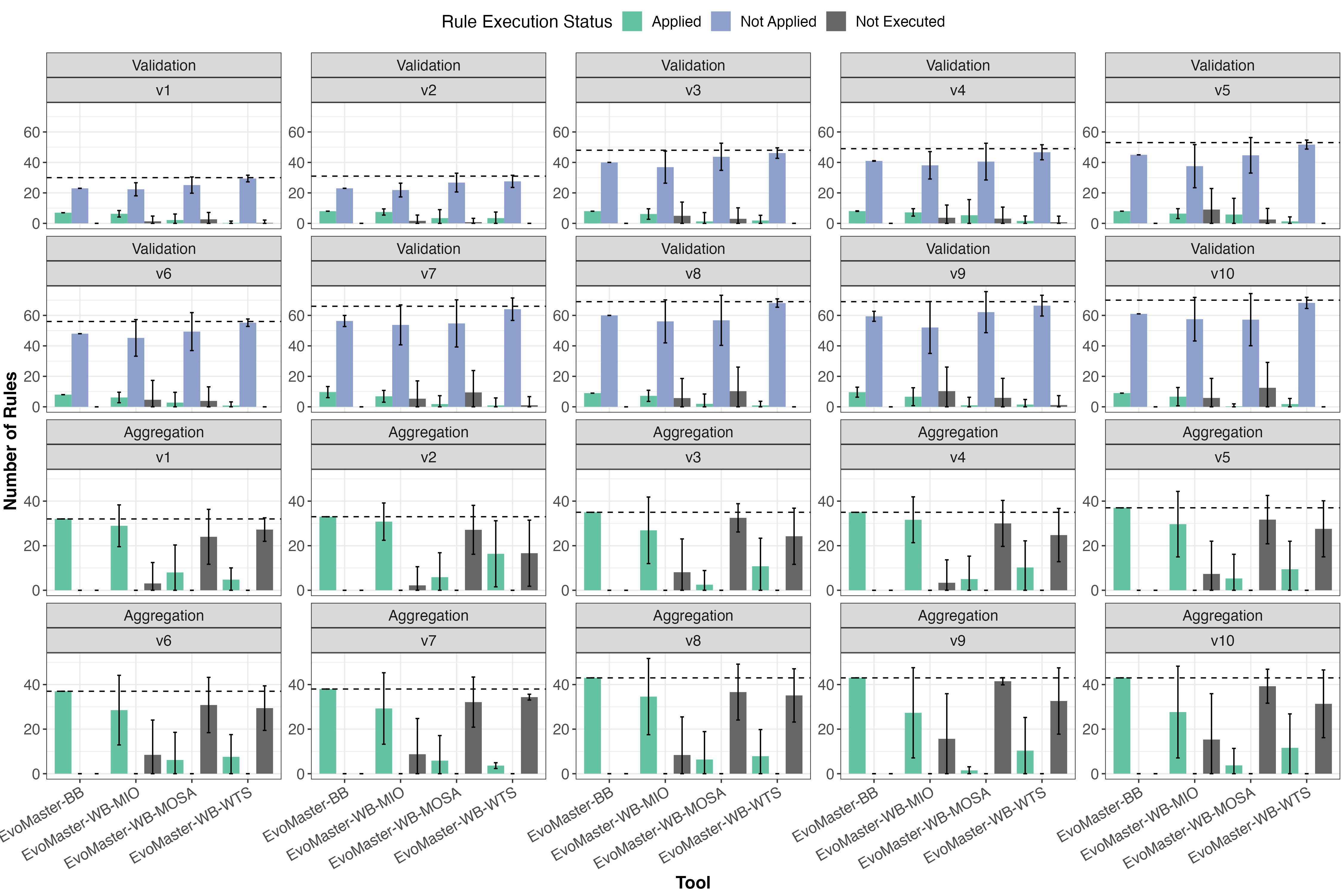}
    \caption{
    Rule execution status for each rule per tool and version across all the repetitions.
    The bars are arithmetic means, the error bars are standard deviations, and the dashed line depicts the number of total rules of a version.
    }
    \label{fig:executed-rules}
\end{figure*}

%% file: src_fig_rule_execution_results.tex
\begin{figure*}[tbp]
    \centering
    \includegraphics[width=\textwidth]{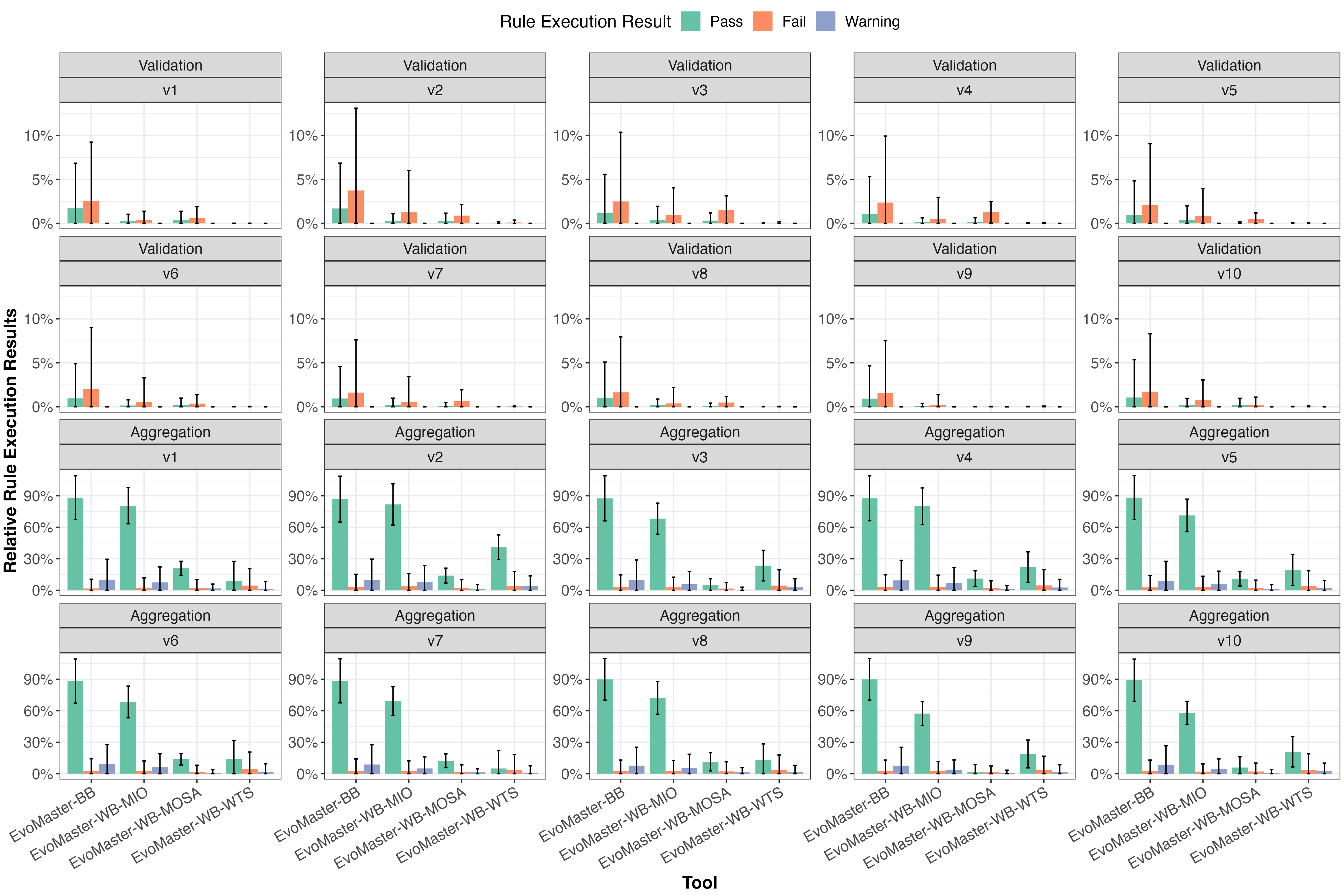}
    \caption{
    Rule execution results relative to the total number of rule executions for each rule, tool, and version.
    The bars are arithmetic means and the error bars are standard deviations.
    }
    \label{fig:rule-execution-results}
\end{figure*}

%% file: src_tab_rule_execution_percentages.tex
\begin{table*}
    \centering
    \caption{
    Rule execution results relative to the total number of rule executions for each rule and tool.
    The values are arithmetic means $\pm$ standard deviations summarized on rule type level across all the individual rules.
    \tool{Production} corresponds to the rule executions from production \guri{}.
    The values for the test generation tools are for the latest version \version{v10}, averaged across all the repetitions.
    }
    \label{tab:rule_execution_precentages}
    
    \include{src_tab_rule_execution_percentages_body.tex}
\end{table*}

%% file: src_tab_rule_execution_percentages_body.tex
\begin{tabular}{lllllll}
\toprule
Rule Type & Tool & \nrcols{5}{Rule Execution Result} \\
\cmidrule(l{1pt}r{1pt}){3-7}
 &  & Pass & Fail & Warning & Not Applied & Not Executed \\
\midrule
Validation & \tool{Production} & \num{26.1918255485415}\%$\pm$\num{37.3160629293796} & \num{0.0535950998398514}\%$\pm$\num{0.153859882156909} & \num{6.55323984204683e-06}\%$\pm$\num{5.52185804086332e-05} & \num{73.7545727983788}\%$\pm$\num{37.3785677226814} & \num{0}\%$\pm$\num{0} \\
 & \evomasterbb{} & \num{0.971985997203439}\%$\pm$\num{3.91636988423105} & \num{1.60856802402935}\%$\pm$\num{6.24635051324596} & \num{0}\%$\pm$\num{0} & \num{97.4194459787672}\%$\pm$\num{7.186198478912} & \num{0}\%$\pm$\num{0} \\
 & \evomasterwbmio{} & \num{0.300524782233319}\%$\pm$\num{0.875876070975904} & \num{1.41167844600728}\%$\pm$\num{3.38094447662236} & \num{0}\%$\pm$\num{0} & \num{93.2877967717594}\%$\pm$\num{6.3102413793449} & \num{5}\%$\pm$\num{5.03610155185335} \\
 & \evomasterwbmosa{} & \num{0}\%$\pm$\num{0} & \num{0}\%$\pm$\num{0} & \num{0}\%$\pm$\num{0} & \num{70.5714285714286}\%$\pm$\num{20.1361823221115} & \num{29.4285714285714}\%$\pm$\num{20.1361823221115} \\
 & \evomasterwbwts{} & \num{0}\%$\pm$\num{0} & \num{0}\%$\pm$\num{0} & \num{0}\%$\pm$\num{0} & \num{100}\%$\pm$\num{0} & \num{0}\%$\pm$\num{0} \\
\midrule
Aggregation & \tool{Production} & \num{99.8835781985487}\%$\pm$\num{0.472240433595079} & \num{0.0188443056469122}\%$\pm$\num{0.0818327448630621} & \num{0.097577495804357}\%$\pm$\num{0.465327357318205} & \num{0}\%$\pm$\num{0} & \num{0}\%$\pm$\num{0} \\
 & \evomasterbb{} & \num{89.0247150069102}\%$\pm$\num{20.2388435928438} & \num{2.38116114429475}\%$\pm$\num{10.6459935557156} & \num{8.59412384879506}\%$\pm$\num{18.2440363557799} & \num{0}\%$\pm$\num{0} & \num{0}\%$\pm$\num{0} \\
 & \evomasterwbmio{} & \num{64.4671507243327}\%$\pm$\num{13.1128592552983} & \num{2.06987267178786}\%$\pm$\num{7.63272003365334} & \num{6.0211161387632}\%$\pm$\num{12.8314522070155} & \num{0}\%$\pm$\num{0} & \num{27.4418604651163}\%$\pm$\num{4.41481448448827} \\
 & \evomasterwbmosa{} & \num{2.77131782945736}\%$\pm$\num{10.6108424666285} & \num{1.91860465116279}\%$\pm$\num{8.58842675914051} & \num{0.658914728682171}\%$\pm$\num{4.32079282609047} & \num{0}\%$\pm$\num{0} & \num{94.6511627906977}\%$\pm$\num{16.3807141552123} \\
 & \evomasterwbwts{} & \num{4.85359759139484}\%$\pm$\num{18.5855288861803} & \num{3.86322979346235}\%$\pm$\num{16.5450927099178} & \num{1.05061447560793}\%$\pm$\num{6.88933983654077} & \num{0}\%$\pm$\num{0} & \num{90.2325581395349}\%$\pm$\num{29.3977948713864} \\
\bottomrule
\end{tabular}

%% file: src_discussion.tex
\section{Discussion and Lessons Learned}
\label{sec:discussion}

This section discusses the results and outlines lessons learned, which provide research opportunities.

\subsection{Need for Domain-Specific Objectives, Targets, and Evaluation Metrics}

From \rqcodecov{} and \rqfaults{}, we see that there are not a lot of differences among the tools;
and from \rqexecutedrules{} and \rqruleresults{}, we observe that the tools can lead to some rules being applied and obtain rule execution results: pass, fail, or warning.
But, it is evident that the current tools do not support testing domain-specific targets well, i.e., they optimize for the \enquote{irrelevant} objectives, e.g., code coverage, and there is much room for improvement.
Going forward, test generation tools require to
\begin{inparaenum}
    \item encode domain-specific objectives in their search, e.g., with added domain-specific search objectives (e.g., rule (result) count or distance metrics to applying rules~\citep{ali:13}), or adding tests that reach unseen domain-specific targets to the archive (similar to \citet{padhye:19a};
    \item keep tests after the search that cover each domain-specific target for regression testing scenarios;
    and
    \item evaluate test generation tools with domain-specific metrics (rule execution status and rule execution results in our case) to show their effectiveness when traditional metrics do not show differences (also discussed by \citet{boehme:22}).
\end{inparaenum}

\subsection{Oracle Problem for Rule Execution Results}

The oracle problem is a well-known problem in software testing~\citep{barr:15b}, which extends also to domain-specific goals.
While we show that the current tools can apply rules with different results, it is unclear if, for a randomly generated test input, a rule is expected to pass, fail, yield a warning, or should not be applied.
Current research does not offer a solution; there simply is no implicit oracle~\citep{barr:15b} for rule executions.
Going forward, this needs to be addressed, and we see three potential aspects:
\begin{inparaenum}
    \item using tests that lead to a specific rule result (pass, fail, warning) and employ them in a regression testing setting, e.g., if a test passes a rule in version \version{1}, it should also pass in version \version{2};
    \item applying differential testing by comparing the outputs of the same random test input to a, e.g., reference implementation for the medical rules, as they should be standardized;
    and
    \item devising metamorphic relations on the rules that are either semantics-preserving (similar to \citet{lu:19}) or are known to lead to invalid rules and comparing the outputs to the correct implementation.
\end{inparaenum}

\subsection{Challenge of Generating Medical Data}

Generating synthetic medical data is a challenge researchers from many fields are trying to tackle~\citep{Goncalves2020, Dankar2021}.
In software testing, generating synthetic and valid medical data is equally important.
The current test generation tools are good at generating syntax-compliant data (e.g., according to an OpenAPI schema definition).
However, the individual variable values (i.e., medical variables) are randomly generated.
This leads to many invalid cancer messages and cases to be checked by the rule engine.
We identify four potential ways to address this challenge in the context of test generation in the future:
\begin{inparaenum}
    \item constrain the valid medical variables through, e.g., the OpenAPI schema definition, either manually or derived from documentation;\footnote{\evomaster{} \version{1.5.0} does not support this; support was added in \version{1.6.0}: \url{https://github.com/EMResearch/EvoMaster/pull/709}}
    \item use generative models such as \glspl{gan} or \glspl{vae} trained on real patient records to generate valid cancer messages and cases to directly test \guri{}~\citep{Hernandez2022};
    \item use generated cancer messages and cases in \evomaster{} either as seeds or when new requests are sampled;
    and
    \item employ \glspl{llm}, possibly trained on \glspl{ehr} or medical text~\citep{singhal:22,lievin:23,yang:22,yunxiang:23}, to generate variable values.
\end{inparaenum}

\subsection{Generality of the Results}

While the results are specific to the case study, i.e., \guri{} at the \gls{crn}, the findings and lessons learned are likely applicable in other contexts.
\begin{inparaenum}
    \item Other countries also have medical registries similar to the \gls{crn}, which also deals with \gls{ehr}.
    These can benefit from the challenges and findings outlined in this paper to introduce automated test generation tools.
    \item Rule-based systems, in general, potentially face similar challenges.
    Once generated tests pass the input validation and execute rules, they will also have to deal with, e.g., domain-specific objectives, the oracle problem, and generating data that executes the rules.
    \item Any system that dynamically loads targets of interest will likely also be affected by the need for domain-specific objectives and targets.
    \item Researchers at a recent Dagstuhl seminar\footnote{\url{https://www.dagstuhl.de/23131} (the report has not been published at the time of writing)} discussed similar challenges, such as domain-specific objectives and targets~\citep{boehme:23:tweet:domain-specific}; comparison to production~\citep{boehme:23:tweet:production}; domain-specific oracles such as reference implementations, differential and metamorphic testing~\citep{boehme:23:tweet:oracles}; and required evaluations that go beyond code coverage and errors~\citep{boehme:23:tweet:evaluation}.
    This shows that while our results are specific to \guri{}, the challenges are important to the research community.
    Consequently, our paper is a valuable case study providing data for these challenges.
\end{inparaenum}

\subsection{Call for Studies with Domain-Specific Goals}

Based on our findings showing that code coverage and uncovered errors are insufficient to evaluate automated test generation tools, we conclude that there is a dire need for more industrial and public sector case studies like ours.
As also outlined by \citet{boehme:22}, code coverage is insufficient to validate test generation tools and fuzzers.
For example, which tool is better when code coverage is equal, or no errors are found?
Exactly this happens in the \gls{crn}'s case.
The research community needs to better understand domain-specific needs for test generation, objectives, and targets to optimize for, and evaluation metrics that are better aligned with stakeholders' interests; and, as a next step, evolve current tools to support these domain-specific needs better.

%% file: src_rw.tex
\section{Related Work}
\label{sec:rw}

\subsection{Test Generation for \gls{rest} \glspl{api}}

Many industrial applications, especially those built with the microservice architecture expose \gls{rest} \glspl{api}.
As a result, there is an increasing demand for automated testing of such \gls{rest} \glspl{api}.
Consequently, we can see a significant rise in publications in recent years~\citep{golmohammadi:22}.
Moreover, several open-source and industrial \gls{rest} \gls{api} testing tools are available such as
\evomaster{}~\citep{arcuri:19,arcuri:18a,arcuri:18b,arcuri:21}, 
RESTler~\citep{atlidakis:19}, 
RestTestGen~\citep{corradini:22},
RESTest~\citep{martin-lopez:21b}, 
Schemathesis~\citep{schemathesis},
Dredd~\citep{dredd},
Tcases~\citep{tcases},
bBOXRT~\citep{laranjeiro:21,evorefuzz},
and
APIFuzzer~\citep{apifuzzer}.
Even though any of these tools can be used in our context, we use \evomaster{}, since it is open-source and has been shown to be the most effective regarding source code coverage and thrown errors among ten different tools in a recent study~\citep{kim:22}.

Generally, \gls{rest} \gls{api} testing approaches are classified into black-box (no source code access) and white-box (requires source code access)~\citep{golmohammadi:22}.
The existing literature has developed testing techniques from three main perspectives to evaluate testing effectiveness~\citep{golmohammadi:22}:
\begin{inparaenum}
    \item coverage criteria, e.g., code coverage (e.g., branch coverage) and schema coverage (e.g., request input parameters);
    \item fault detection, e.g., service errors (i.e, \gls{http} status code 5XX) and \gls{rest} \gls{api} schema violations;
    and
    \item performance metrics, i.e., related to the response time of \gls{rest} \gls{api} requests.
\end{inparaenum}
To achieve these objectives, various algorithms have been developed in the literature.
For instance, \evomaster{} has implemented several \glspl{ea} including random testing~\citep{arcuri:19,arcuri:18b,arcuri:21}.
Various extensions have been also proposed to \evomaster, such as handling sequences of \gls{rest} \gls{api} calls and their dependencies~\citep{zhang:21b}, handling database access through \gls{sql}~\citep{zhang:21a}, and testing \gls{rpc}-based \glspl{api}~\citep{zhang:23}.
In our case, we have access to the source code of \guri{}; therefore, we employ both the black-box and the white-box (parameterized with three \glspl{ea}) tools of \evomaster{}.
However, an extended investigation in the future may include other tools.

Compared to the literature, our main contribution is applying an open-source \gls{rest} \gls{api} testing tool in the real-world context of the \gls{crn}.
We assess the tool's effectiveness in achieving code coverage, errors found, and domain-specific metrics (e.g., related to medical rules defined for validating and aggregating cancer messages and cancer cases).

\subsection{Development and Testing of Cancer Registry Systems}
In our recent paper with the \gls{crn}~\citep{laaber:23b}, we assess the current state of practice and identify challenges (e.g., test automation, testing evolution, and testing \gls{ml} algorithms) when testing \gls{crn}['s] \gls{caress}.
This paper is the first concrete step towards handling those challenges, particularly assessing the effectiveness of an existing testing tool in the \gls{crn}'s context.
Two other recent works build cyber-cyber physical twins for \guri{}~\citep{lu:23} and incorporate \gls{ml} classifiers into \evomaster{} to reduce testing cost~\citep{isaku:23}.

In the past, we developed a model-based engineering framework to support \gls{caress} at the \gls{crn}~\citep{wang:16b}.
The framework aims to create high-level and abstract models to capture various rules, their validation, selection, and aggregation.
The framework is implemented based on the \gls{uml} and \gls{ocl}, where the \gls{uml} is used to capture domain concepts and the \gls{ocl} is used to specify medical rules. 
The implementation of the framework has been incorporated inside \guri{}, which is the subject of testing in this paper.
As a follow-up, we also developed an impact analysis approach focusing on capturing changes in rules and assessing their impact to facilitate a systematic evolution of rules~\citep{wang:17}.
Finally, we also developed a search-based approach to refactor such rules regarding their understandability and maintainability~\citep{lu:19}.
Compared to the existing works, this paper focuses on testing \guri{} as a first step toward building cost-effective testing techniques at the \gls{crn}.

%% file: src_conclusions.tex
\section{Conclusions}
\label{sec:conclusions}
\glsresetall{}

This paper reports on an empirical study evaluating the test effectiveness of \evomaster{}'s four testing tools on a real-world system, the \gls{crn}'s \gls{caress}.
\Gls{caress} is a complex software system that collects and processes cancer patients' data in Norway and produces statistics and data for its end users.
Our results show that all the studied testing tools preform similarly regarding code coverage and errors reported across all the studied versions.
However, in terms of domain-specific metrics, \evomaster{}'s black-box tool is more effective; hence, we recommend it for the \gls{crn} as a starting point.
We also provide lessons learned that are beneficial for researchers and practitioners.